\documentclass[12pt]{article}

\usepackage{amsmath,amssymb,epsfig}
\usepackage{wrapfig}
\usepackage{floatflt}
\usepackage{graphics}

\newcommand{\be}{\begin{equation}}
\newcommand{\ba}{\begin{eqnarray}}
\newcommand{\ea}{\end{eqnarray}}

\def\d{\delta}

\def\l{\lambda}
\def\m{\mu}

\def\G{\Gamma}

\def\ca{{\cal A}}
\def\cb{{\cal B}}

\def\ch{{\cal H}}

\def\cm{{\cal M}}

\def\cz{{\cal Z}}

\newcommand{\bbC}{{\Bbb C}}
\newcommand{\bbH}{{\Bbb H}}

\newcommand{\cL}{{\cal L}}

\fontfamily{yfrak}

\begin{document}

\vskip 15mm

\begin{center}

{\Large\bfseries Intersecting Connes Noncommutative Geometry with Quantum Gravity  
}

\vskip 4ex

Johannes \textsc{Aastrup}$\,^{1}$ \&
Jesper M\o ller \textsc{Grimstrup}\,$^{2}$ 

\vskip 3ex  

$^{1}\,$\textit{Institut f\"ur Analysis, Universit\"at Hannover \\
  Welfengarten 1, D-30167 Hannover, Germany}
\\
E-mail:  
\texttt{aastrup@math.uni-hannover.de}
\\[3ex]
$^{2}\,$\textit{NORDITA \\Blegdamsvej 17, DK-2100 Copenhagen, Denmark}
\\
E-mail: \texttt{grimstrup@nordita.dk}

\end{center}

\vskip 5ex

\begin{abstract}

An intersection of Noncommutative Geometry and Loop Quantum Gravity is proposed.
Alain Connes' Noncommutative Geometry provides a framework in which the Standard Model of
particle physics coupled to general relativity is formulated as a unified, gravitational
theory. However, to this day no quantization procedure compatible with this framework
is known. In this paper we consider the
noncommutative algebra of holonomy loops on a functional space of certain
spin-connections. The construction of a spectral triple is outlined and ideas
on interpretation and classical limit are presented.

\end{abstract}

\newpage




\section{Introduction}

 The framework of noncommutative geometry \cite{ConnesBook} suggests an appealing solution to
  one of the central riddles of theoretical physics, the unification of
  general relativity and the standard model. However, the
  noncommutative formulation of the standard model is intrinsically classical
  and no notion of quantization within this framework is known. In this paper
  we attempt to address this problem by suggesting an intersection of
  noncommutative geometry with principles of Loop Quantum Gravity (LQG): The
  idea presented is to apply the machinery of noncommutative geometry to the
  algebra of holonomy loops. This algebra is naturally noncommutative and stores topological information about an underlying space of
  connections. The goal is a spectral triple over this
  functional space of geometries. \\

In the following we will outline and clarify ideas already presented in
\cite{Aastrup:2005yk}. Emphasis is put on the general idea rather than
technical details. First,
in section \ref{sec1}, we
briefly introduce the standard model framed within noncommutative geometry and propose an application of noncommutative geometry to a
functional space of Euclidean gravity. In section \ref{sec2} we outline the
construction of a spectral triple and, in section \ref{sec3}, an
interpretation of the construction is presented. Finally we conclude
and discuss various problems in section \ref{sec4}.

\section{Noncommutative geometry, the standard model and quantization}
\label{sec1}

A noncommutative geometry in the sense of Connes is determined by a spectral
triple $(\cb,D,\ch)$ which consist of a $\ast$-algebra $\cb$
represented on a Hilbert space $\ch$ on which a self-adjoint
unbounded operator $D$, the Dirac operator, acts. The triple is normally required to
satisfy a set of seven axioms proposed by Connes \cite{Connes:1996gi}.
Ordinary Riemannian
spin-geometries form a subset in this framework and are described by
commutative $C^\ast$-algebras. Here, the underlying manifold $\cm$ emerges as the
spectrum of the algebra and a differential structure is provided by the Dirac operator.
For example, the distance between points $x,y\in\cm$ can be
formulated algebraically \cite{Connes:1996gi}
\ba
d(x,y) = \sup_{f\in\cb}\Big\{|\chi_x(f) -\chi_y(f)| \Big| \|[D,f]  \|\leq 1  \Big\}\;,
\label{distanc}
\ea
where $\|\cdot\|$ on the rhs is the supremum norm and $\chi_x$ is the character
corresponding to the point $x$, i.e. $\chi_x(f)=f(x)$. 
Further, we can recover the Clifford algebra, and so differential forms, by
considering commutators of the
Dirac operator
\ba
\xi = f_1 [D,f_2]\ldots [D,f_k]\;,\quad f_i\in\cb\;.
\label{forms}
\ea
Differential structures
such as (\ref{distanc}) and (\ref{forms})
continue to make sense even when the algebra is noncommutative. Such algebras
can not always be identified as function algebras over manifolds and the set of
points, the spectrum of the algebra, is often reduced and may, although the algebra is infinite dimensional, be
discrete. However, noncommutative geometry permits differential structures
which treats this broad variety of spaces on an equal footing.\\

\noindent {\bf The Standard Model}\\

A special class of noncommutative geometries are the {\it almost commutative
  geometries} described by spectral triples of the form
\ba
\cb&=&C^\infty(\cm)\otimes \cb_F\;, 
\nonumber\\
D&=&\not\hspace{-1.5mm} D \otimes \gamma^{d} +
1\otimes D_F\;,
\nonumber\\ 
\ch&=& L^2(\cm,S\times M_n(\bbC))\;,
\label{almost}
\ea
where $\cb_F$ is a finite dimensional matrix algebra, $\not\hspace{-1.5mm} D$ is
the usual space-time Dirac operator and $D_F$ is a
matrix operator satisfying certain criteria. The dimension of the (Riemannian,
compact) manifold $\cm is $d-1$ $and $\gamma^{d}=\gamma^1\cdot\ldots\cdot\gamma^{d-1}$ are the
  gamma matrices. The spinor-bundle is denoted by $S$. The
  almost commutative geometry given by 
\ba
\cb_F = \bbC\oplus \bbH\oplus M_3(\bbC)\;,
\label{SMalgebra}
\ea
where $\bbH$ denotes quarternions, is the algebra which forms the basis of
the formulation of the standard model in terms of noncommutative
geometry \cite{Connes:1990qp,Connes:1996gi}. Here the
Hilbert space is the Hilbert space of the entire fermionic content of the
standard model and $D_F$ contains information about
the fermion masses.
It is natural to consider fluctuations around the Dirac operator of the form
\ba
D\rightarrow \tilde{D} = D + A + J A J^\dagger\;,
\label{unitaryfluctuation}
\ea
where $J$ denotes charge conjugation and 
\ba
A = \sum_i a_i [D,b_i]\;,\quad a_i,b_i\in\cb\;,
\label{oneform}
\ea
is a general one-form in
the sense of (\ref{forms}). 
Combining
(\ref{oneform}) with (\ref{SMalgebra}) one derives the entire bosonic
sector of the standard model with the Higgs boson
emerging as an integrated part of the noncommutative gauge field.

The
algebraic equivalence of diffeomorphism invariance is invariance under automorphisms of the
algebra.
For noncommutative algebras like (\ref{almost}) the
automorphism group is
larger than the diffeomorphism group, including also inner automorphisms, i.e.
gauge transformations.
Requiring invariance under automorphisms of the algebra leads to the
spectral action principle which states that physics only depends
on the spectrum of the Dirac operator. 
This statement is quantified by the action formula \cite{Chamseddine:1991qh,Chamseddine:1996rw}
\ba
S=
\langle\xi |
\tilde{D} |\xi \rangle + \mbox{Trace} \;\varphi\left( \frac{\tilde{D}}{\Lambda}\right)\;,
\label{acfor}
\ea
where $\xi$ is
a Hilbert state and $\varphi$ a suitable function selecting eigenvalues
below the cutoff $\Lambda$. Equation (\ref{acfor}) contains the entire classical action of the
standard model coupled to gravity. 

It is intriguing that Yang-Mills-Higgs models compatible with Connes
noncommutative geometry appear to be rather rare \cite{Jureit:2005tw,Schucker:2005aa,Jureit:2005ay,Iochum:2003xy}.

To conclude, the standard model coupled to gravity can be formulated as pure
geometry over a 'space' characterized by an almost commutative
algebra. Fermions are linked to the metric structure and the bosonic sector arises through fluctuations around the Dirac operator, the 'metric' of noncommutative geometry.\\

\noindent{\bf Quantization}\\

The fact that the standard model can be
formulated as a gravitational theory suggests that it
can not be quantized in any straightforward manner within the framework of noncommutative geometry since such a
quantization would, accordingly, involve quantum gravity. 
On the other hand, this implies that the search for a suitable
quantization scheme might pass through quantum gravity. 
This is the problem which we wish to address in this paper: How principles of noncommutative
geometry could be unified with ideas in quantum gravity.

\subsection{Intersecting noncommutative geometry and quantum gravity}

Consider first quantum field theory. It involves, via Feynman path integrals, integration theory
over spaces of field configurations. The central object is the partition
function, the generating functional for Greens functions
\[
\cz[J] = \int [d\Phi]\exp(-\tfrac{\rm{i}}{\hbar}S[\Phi,J])\;,
\]
where $\Phi$ denotes the field content of the model characterized by the
classical action $S[\Phi]$ coupled to external fields $J$.
We now propose the following: Since Connes formulation of the standard model lacks a clear quantization
procedure and since quantum field theory deals with integration theory over
spaces of field configurations, it seems natural to try to apply the machinery of
noncommutative geometry to functional spaces. Further, since Connes formulation of the standard model is essentially a
gravitational theory we suggest investigating a configuration space related
to gravity.

In fact, a configuration space suitable for our purposes has already been investigated in the
literature. Loop Quantum Gravity (LQG) \cite{Ashtekar:2004eh,Rovelli:2004tv,Smolin:2004sx} is an attempt to quantize general
relativity using methods of canonical quantization. The configuration space relevant for LQG is a space ${\ca}$ of $SU(2)$
connections which are interpreted as certain spin-connections living on a
3-dimensional hyper-surface. This surface emerges from a foliation of
4-dimensional space-time which is needed for the quantization
procedure.

Central to LQG is an algebra of Wilson loops $W(L)$
which form an abelian algebra of observables on the space of connections
\ba
W(L): {\ca}&\rightarrow& \bbC\;,\
\nonumber\\
\nabla &\rightarrow & Tr\; \mbox{Hol}(L,\nabla)\;,
\ea
where $\mbox{Hol}(L,\nabla)$ is the holonomy of the connection $\nabla$ along the
loop $L$ and $Tr$ is the trace with respect to the representation of the
group. One of the advantages of this formulation is that it permits a natural
implementation of diffeomorphism invariance in a way that leads to a countable
structure, including a
separable Hilbert \cite{Fairbairn:2004qe}: Roughly, the set of Wilson loops form certain
labeled, oriented
graphs of increasing complexity and, up to diffeomorphisms, only the
structure of graphs is relevant. This structure is countable\footnote{In fact, it turns out that so-called extended
diffeomorphisms are required to obtain countable structures. Extended
diffeomorphisms permit finitely many non-smooth points.}.

We believe that there exist a natural intersection between LQG and
noncommutative geometry: instead of using Wilson loops we suggest to study the noncommutative algebra of
holonomy loops themselves. By avoiding the trace the gauge symmetry of local Lorentz
transformations is preserved. Further, since the objective is to apply the machinery of
noncommutative geometry to the functional space, rather than a canonical
quantization procedure, we propose to consider space-time as a whole and avoid
a foliation. Thus, we consider an algebra of space-time holonomy loops
\ba
L: {\ca}&\rightarrow & G\;,
\nonumber\\
\nabla&\rightarrow& \mbox{Hol}(L,\nabla)\;,
\ea
where $G$ is the symmetry group. Since compactness of the gauge group is at present
needed for the analysis, we are at first limited
to consider Euclidean gravity with, for example,
\[
G= SO(4)\;.
\]
Finally, rather than postulating constraints on the Hilbert space, such as the
Hamilton constraint in LQG, we suggest to apply the spectral action principle
\cite{Chamseddine:1991qh,Chamseddine:1996rw}. This amounts to seek
physical information in the spectrum of the Dirac operator.

This intersection of LQG and noncommutative geometry contains all the ingredients we are looking
for: Integration theory over a functional space related to
gravity\footnote{Since the algebra of holonomy loops lives on a space of
  connections the corresponding Hilbert space will be equiped with an inner
  product involving a functional integration \cite{Aastrup:2005yk}.} involving a natural 
noncommutative algebra.

The holonomy loops are matrix valued and can, as we will show below, be heuristically argued to entail an almost
commutative algebra in a classical limit characterized by a single
space-time geometry, that is, a single point in ${\ca}$. This is encouraging and
provides us with the hope that low energy physics characterized
by an almost commutative algebra may arise as the classical limit of a pure quantum gravity.

\section{The construction}
\label{sec2}

Let us go into some details. As already mentioned, the space of interest is the space $\ca$ of connections with values in the Lie-algebra of a compact group $G$. As algebra of observables it is natural to take functions on $\ca$. A natural collection of functions on $\ca$ is the traced holonomies i.e. given a closed loop $L$ the associated function is
$$Tr\circ f_L:\ca \ni \nabla \to Tr(Hol(L,\nabla))\;.$$  
It can be shown that this is a complete set of functions,
i.e. that the algebra of linear combinations of these functions completely
determines $\ca$ modulo gauge equivalence \cite{Ashtekar:1993wf}.

Since we are interested in a noncommutative algebra we will take untraced holonomies and therefore  get functions over $\ca$ with values in the matrix representation of $G$. The algebra we therefore want to consider is the algebra of all linear combinations of such functions
i.e. functions on the form 
$$a_1f_{L_1}+\ldots +a_nf_{L_n}\;.$$
This algebra comes with a norm, namely the usual sup-norm over $\ca$. The
completion in this norm will be a $C^*$-algebra. 

Similar to LQG  this $C^*$-algebra can be seen as matrix valued functions over a space $\bar {\ca}$ containing $\ca$ as a dense subset and $\bar{\ca}$ can be written as $Hom (\cL_{x_0} ,G)$ where $\cL_{x_0}$ is the hoop group based in a given point $x_0$, i.e. loops modulo trivial backtracking and reparameterization, the group structure being composition of loops.

\subsection{Inductive systems and geometrical structures}

\begin{figure}
\centering
\includegraphics{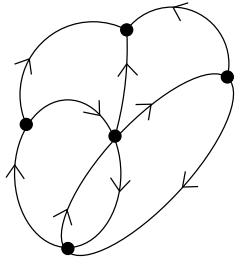}
\caption{A generic graph.}
\end{figure}
Construction of geometrical structures directly on $\ca$ does not seem
easy. Instead we will start by looking at $\ca$ as "seen from" a finite
collection of loops, $L_1,\ldots,L_n$. This can be interpreted as a
regularization of the functional space $\ca$. Seen from this collection of loops $\ca$ just looks like $G^n$: Namely, a connection $\nabla \in \ca$  gives rise to an element in $G^n$ via 
$$(Hol(L_1,\nabla ),\ldots , Hol(L_n,\nabla ))\;.$$ 
Therefore, in this regularized picture the {\it functional space} $\ca$ can be identified with a
  manifold\footnote{This makes $\ca$, or rather its closure $\bar{\ca}$, a so-called
    {\it pro-manifold} since it can be identified with a projective system of
    manifolds. This leaves $\bar{\ca}$ with nice properties. In particular,
    $\bar{\ca}$ has a canonical topology \cite{Marolf:1994cj}.}. Thus, a connection is a point
  $(g_1,\ldots,g_n)$ on $G^n$. Each value $g_i$ should be thought of as the
  holonomy along the i'th loop, $L_i$. On $G^n$ it is easy to construct various structures. If we for example want to
construct a Hilbert space, it is natural to take $L^2(G^n)$, the space of
square integrable functions on $G^n$ with respect to the Haar measure on
$G^n$. These are functions on the connections restricted to the graph, denoted
$\Gamma$, spanned by $ L_1, \ldots , L_n$. The inner product on this Hilbert space should be interpreted as a
functional integral over connections and any derivations acting on this space
as functional derivations. In general, loops may intersect and
the corresponding graph will consist of a number of edges and vertices, see
fig.1. In this case the number of edges, $n$, corresponds to the number of
independent line segments and the corresponding space will be $G^n$ (see \cite{Ashtekar:1993wf} or
\cite{Ashtekar:1994wa} for details).
 If we have another collection of loops $ L'_1,\ldots , L_m'$ whose graph $\Gamma'$ contains the graph $\Gamma$ the functions living on the graph $\Gamma$ can be considered as functions living on $\Gamma'$. Formally we get an embedding of Hilbert spaces $L^2(G^n)$ into $L^2(G^m)$, see fig.2.

The construction of the space of square integrable functions on $\ca$ follows by considering functions that lies in one of the spaces $L^2(G^n)$ associated to a finite collection of loops and identify them if they coincide on some graph containing both of them. In this way we get functions on connections restricted to all graphs, i.e. functions on connections on the entire $M$, that is $\ca$.

The formal language of the construction is that of inductive limits of Hilbert spaces.

\subsection{The spectral triple}

\begin{figure}
\centering
\includegraphics{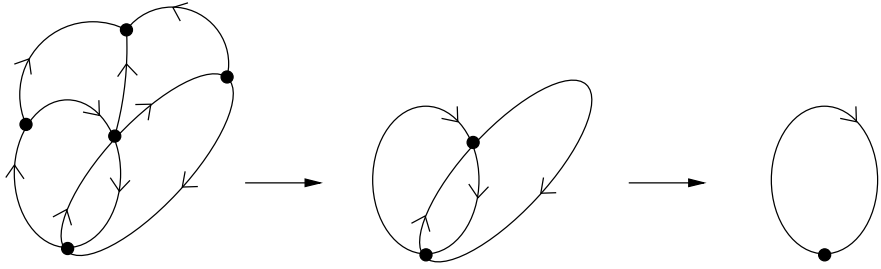}
\caption{Part of a projective system of graphs.}
\end{figure}
The aim is to construct a spectral triple on the algebra of untraced holonomies. For a finite collection of loops $L_1,\ldots, L_n$ there is an obvious Hilbert space\footnote{We would prefer the Hilbert space $L^2(G^n,S)$ involving spin structure on $G^n$. However, this entails embedding problems for which we have found no solution, see \cite{Aastrup:2005yk}.} and Dirac operator, namely 
$L^2(G^n,Cl(TG^n))$ and the usual Euler-Dirac operator $D=d + d^*$. By
applying the same construction as in the previous section we get the Hilbert
space $L^2(\bar{\ca} , Cl(T\bar{\ca} ))$ and an Euler-Dirac operator $D$ on
this space. Here $Cl$ denotes the Clifford algebra. By tensoring with $M_N$, $N$ being the size of the representation of $G$, we end up with a triple
\ba
(A_{ut},D\otimes 1,L^2(\bar{\ca},Cl(T\bar{\ca})\otimes M_N))\;,
\label{tripleA}
\ea
where the algebra $A_{ut}$ of untraced holonomies acts pointwise on the
$M_N$-part of $Cl(T\bar{\ca})\otimes M_N$. This is our spectral triple. 

The triple (\ref{tripleA}) is however far from fulfilling the conditions of Connes, since the
kernel of $D$ is infinite and not even separable (recall that the Hilbert
space is not separable). In the next section we will
explain how the symmetry group of diffeomorphisms can be used to obtain
countable structures.

Several problems arise during the construction of the triple. The key question is to
construct an embedding of Hilbert spaces $L^2(G^n,Cl(TG^n))$ into
$L^2(G^m,Cl(TG^m))$ compatible with the embedding of the corresponding
graphs. This boils down to the construction of a metric on $G^n$ compatible
with the embeddings. In \cite{Aastrup:2005yk} we circumvented the problem by discarding
intersecting loops. This, however, is clearly unsatisfactory since the
inclusion of intersecting loops is essential in order to obtain the correct
projective limit. We will address this problem in
a forthcoming publication.\\

\subsection{An $n=3$ graph}

\begin{figure}
\centering
\includegraphics{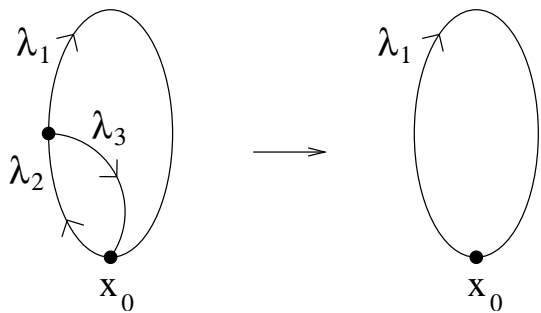}
\caption{An $n=3$ graphs projected down to the $n=1$ graph.}
 \label{F}
 \end{figure}
Before we continue with diffeomorphism invariance let us use an example to
clarify the construction and point out its weak point.
Consider the graph $\G$ consisting of two vertices and three edges as shown on the lhs of
fig \ref{F}. The space $\ca$ restricted to this graph is identified with the manifold   
\[
\ca_\G\simeq G^3\;,
\]
which means that a connection is characterized by three group elements,
$\nabla=(g_1,g_2,g_3)$. Each group element $g_i$ is interpreted as the
parallel transport of $\nabla$ along the line segment $\l_i$. We are interested in the noncommutative algebra
of (holonomy-) loops which in this case is generated by the elements
\ba
L_{12}=g_1\cdot g_2\;,\;\; L_{13}=g_1\cdot g_3\;,\;\; L_{3^{*}2}=g_3^{-1}\cdot g_2\;.
\label{algebra}
\ea
since combined line-segments such as $\l_1\circ\l_2$ form closed loops. 

Next, let $\{e_i \}$ be a global basis of the tangent bundle to $G^3$ and
$\{\bar{e}_i \}$ its dual. The Clifford bundle is constructed by imposing the anti-commutator
relation $\{\bar{e}_i,\bar{e}_j\}=2\langle \bar{e}_i|\bar{e}_j\rangle$.  We construct the Hilbert space
\ba
\ch_\G = L^2(G^3,Cl(T^\ast G^3)\otimes M_N(\bbC))\;,
\label{hrum}
\ea
of matrix-valued functions on $G^3$ with additional values in the Clifford
bundle. In (\ref{hrum}) $N$ is the size of the representation of $G$. The algebra of loops (\ref{algebra}) acts by matrix multiplication on
the factor $M_N(\bbC)$ in (\ref{hrum}). For example, the loop $L_{12}$ acts on $\ch_\G$ by
\[
L_{12}\Psi(\nabla):= (id\otimes\nabla(L_{12}))\Psi(\nabla)= (id \otimes g_1\cdot g_2) \Psi(g_1,g_2,g_3)\;,\quad\Psi\in\ch_\G\;.
\]
The inner product consist of three components: Integration over the group,
trace of the matrices and the inner product on the Clifford bundle. Thus, if
we write $\Psi=\Psi_{ij}e_{k_1} \cdots e_{k_l}$ where $i,j$ are indices of the
matrix the norm associated to the inner product reads (sum over repeated indices)
\[
\langle\Psi^1|\Psi^2\rangle = \int d\m\; \Psi^1_{ij}
(\Psi^2_{ij})^\ast\langle e^1_{k_1} \cdots e^1_{k_l} |e^2_{k_1} \cdots
e^2_{k_l} \rangle \;,\quad \Psi^i\in\ch_\G\;\forall i\in\{1,2\}\;,
\]
where $d\m$ is the Haar measure on $G^3$ and $(\Psi_{ij})^\ast$ is the complex
conjugate of $\Psi_{ij}$. Notice that by taking the matrix
trace we turn holonomy loops into gauge-invariant Wilson
loops. Further, the integration over the group should, as already mentioned, be interpreted as a
functional integral since each point in $G^3$ represents a connection; more on
this later. 

The Dirac operator has the form
\ba
D = \sum_i \bar{e}_i\nabla_{e_i}\;,
\label{diraC}
\ea
where $\nabla_{e_i}$ denotes the Levi-Civita connection in the direction of $e_i$.

As shown in fig. \ref{F} the graph $\G$ is related to the 
graph of a single loop via the projection
\ba
P(g_1,g_2,g_3)=g_1 \cdot g_2\;.
\label{ProJ}
\ea
So far we did not specify the inner product on $T^\ast G^3$. In fact, this
turns out to be a crucial point. Let us demonstrate this by taking the
simplest case, $G=U(1)$. The most obvious attempt for an inner product would
be the product metric on $G^3$; i.e. the three copies of $G$ are
orthogonal. This is however not going to work for the projection (\ref{ProJ}) for the
following reason:
We use the coordinate $\theta$ on $U(1)$ given by $g=\exp(2 \pi i\theta)$. In
such coordinates the projection (\ref{ProJ}) is given by
\[
P(\theta_1,\theta_2,\theta_3)=\theta_1+\theta_2 :=\theta\;.
\]
Denote by $d\theta_i$ an orthonormal basis for the cotangent bundle $T^\ast G^3$. The inner
product on the cotangent bundle is given by
\[
\langle d\theta_i |d\theta_j \rangle = \d_{ij}\;.
\]
We consider now the norm of the vector $d\theta_1 + d\theta_2\in T^\ast G^3$
\ba
\langle d\theta_1+ d\theta_2|d\theta_1+ d\theta_2 \rangle = 2\;.
\label{no1}
\ea
On the other hand, the norm of the vector $d\theta\in T^\ast G$ is
\ba
\langle d\theta | d\theta\rangle =1\;.
\label{no2}
\ea
However, the two vectors $d\theta_1+d\theta_2$ and $d\theta$ are related by
the induced projection $P^\ast$
\[
P^\ast: T^\ast G \rightarrow T^\ast G^3\;,\quad P^\ast (d\theta)=d\theta_1+d\theta_2\;,
\]
which means that the inner product on the cotangent spaces is not compatible
with the projection (\ref{ProJ}) since (\ref{no1}) does not equal
(\ref{no2}). Thus, the Hilbert space (\ref{hrum}) is not compatible
with the projection either\footnote{These difficulties were encountered also
  in \cite{Ashtekar:1994wa} where the authors suggested a metric compatible with all
  projections. However, the metric constructed turns out to be degenerate.}. This means that we cannot take the
inductive limit of the spectral triple candidate (\ref{algebra})+(\ref{hrum})+(\ref{diraC}) in any consistent way.

We see two obvious strategies to solve this problem. First, one can
attempt to construct an inner product on the cotangent space which is
compatible with projections of the form (\ref{ProJ}). Such an inner product must
necessarily leave different copies of $G$ non-orthogonal, i.e.
\[
\langle d\theta_1 |d\theta_2 \rangle\not= 0\;.
\]
This will change the Dirac operator since it contains the metric.

 Second, one can attempt to construct a projective system which avoids
projections of the form (\ref{ProJ}) altogether.

We shall not elaborate further on these difficulties here as we shall address
them elsewhere.

\subsection{Diffeomorphism invariance}

\begin{figure}
\centering
\includegraphics{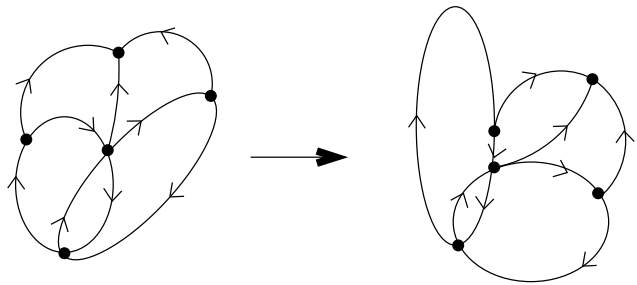}
\caption{Two graphs related by a diffeomorphism.}
 \end{figure}

We will now address the problem of diffeomorphism invariance (see
\cite{Aastrup:2005yk} for details). The naive idea is to define the Hilbert space $\ch_{Diff}$ of
diffeomorphism invariant states  by
\begin{equation} \label{dif}
\sum_{\phi \in Diff (M)} \phi(\xi), \quad \xi \in L^2(\bar{\ca},Cl(T\bar{\ca})\otimes M_N)).
\end{equation}
This sum is manifestly diffeomorphism invariant, and of course does not make sense. However, the space $L^2(\bar{\ca},Cl(T\bar{\ca})\otimes
M_N))$ is made up of functions living on graphs. If we consider a graph, it can be deformed, by a diffeomorphism, into
any other graph with the same combinatorial structure (or at least by an extended diffeomorphism, see \cite{Fairbairn:2004qe}). As the part of
$\ch_{Diff}$ living on graphs with a given combinatorial structure we can therefore just take the Hilbert space of functions
living on one fixed graph with this combinatorial structure. This however, would be to overlook the diffeomorphisms mapping the
graph into itself. Hence as the part of $\ch_{Diff}$ living on graphs with a given combinatorial structure we take functions of
the form (\ref{dif}), where $\xi$ belongs to the Hilbert space of a fixed graph with the given combinatorial structure. This time the
sum makes sense, since it is finite (it basically consist of permutations of edges and we also strictly speaking need to weight
the sum).

We thus have the parts of $\ch_{Diff}$ corresponding to each combinatorial structure a graph can have. The construction of
$\ch_{Diff}$ is then the same as for $L^2(\bar{\ca} ,Cl(T\bar{\ca})\otimes M_N))$ by considering embeddings of smaller graphs into bigger
graphs. In this way we get a Hilbert space corresponding to one infinite "graph" containing only the combinatorics of graphs on
the manifold.

The construction of the algebra is similar and the Dirac operator on the space
of connections descends to $\ch_{Diff}$ since it is diffeomorphism invariant.

The Hilbert space $\ch_{Diff}$ is separable, since the combinatorics
of graphs is countable.\\

It is not clear to us whether it is wise to treat diffeomorphisms as
described above, or whether one should keep them as elements of the
automorphism group. In any case, we find it intriguing that, up to (extended)
diffeomorphisms, only countable structures remain.

\section{Discussion}
\label{sec3}

Let us here discuss the interpretation of the setup: \\

\noindent{\bf The inner product as a path-integral}\\

In the previous section we mentioned the Hilbert space of states identified up to
diffeomorphisms. Let $\Psi\in\ch_{Diff}$ and consider automorphism
invariant quantities of the form
\ba
\langle \Psi | \ldots | \Psi\rangle\;.
\label{inner-integration}
\ea
The inner product involves, after the appropriate limit is
taken, an integration over $G^\infty$ which is, up to diffeomorphisms, associated to the space of
connections ${\ca}$. Thus, the interpretation of (\ref{inner-integration})
as a path integral of the form
\ba
\sim\int_{{\ca}/diff}\ldots
\label{int-int}
\ea
lies at hand. The object in (\ref{int-int}) involves an integration which can
be interpreted as a sum over all
'geometries' up to diffeomorphisms. One could impose a spectral action
principle \cite{Chamseddine:1991qh} and consider
automorphism invariant quantities of the form\footnote{such objects remain to
  be defined rigorously.} 
\[
\langle \Psi | D | \Psi\rangle\;,\quad Tr D^2\;,\quad \ldots
\]
which might combine to some sort of an effective action of quantum gravity.\\

\noindent {\bf An emerging almost commutative algebra}\\

Consider a classical connection $\nabla_0$ in ${\ca}$. In a classical limit a state
$\Psi(\nabla)$ should peak around a classical geometry, for example
$\nabla_0$. In such a limit the loop algebra acts like
\[
L \cdot\Psi(\nabla) = \nabla(L)\cdot \Psi(\nabla) = \nabla_0(L)\cdot \Psi(\nabla)\;.
\]
Since each loop $L$ will generate an element $\nabla_0(L)$ in $G$ the entire
algebra of loops will reduce to a (sub) matrix algebra $M_N(\bbC)$. In case $\nabla_0$ equals the flat geometry, $\nabla_0(L)\equiv I$,
the algebra appears to be abelian, simply $\bbC$.
Further, it seems
natural to 'average' the construction over the whole manifold since the
choice of a basepoint $x_0$ breaks part of the diffeomorphism invariance. This
amounts to multiplying the matrix algebra with the function algebra of the manifold. The
result is an almost commutative algebra
\ba
C^\infty(\cm)\otimes M_n(\bbC)\;,
\label{formi}
\ea
or a sub-algebra thereof. With the connection $\nabla_0$ we are
provided with a differential structure on the manifold and thus with a Dirac
operator and a corresponding Hilbert space. Thus, these heuristic arguments show that an almost
commutative algebra may emerge in the classical limit and that the matrix part
of the algebra is related to the group algebra of the Lorentz group. 

Recall that an algebra
of the general form (\ref{formi}) provides, within the framework of noncommutative geometry, the basis for a Yang-Mills-Higgs model
coupled to gravity. This is encouraging and provides
us with the hope that low-energy physics may be recovered in the classical
limit of pure quantum gravity. 

The interesting question of what a
semiclassical state might look like remains.\\

\noindent{\bf An interpretation of the Dirac operator}\\

A connection is determined by holonomies along loops. In the projective
system described here we consider first a finite number of loops and a
connection is thus described 'coarse-grained' by assigning group elements to each of the
finitely many elementary
loops (or edges in the corresponding graph). The Dirac operator takes the derivative on each of these copies of the
group $G$ and throws it into the Clifford bundle. In this way the Dirac
operator resembles a functional derivation operator.  

We interpret this Dirac operator as intrinsically 'quantum'
since it bears some resemblance to a canonical conjugate of the
connection. Heuristically, we write
\ba
D \sim\frac{\d}{\d\nabla}
\label{heu1}
\ea
and 
\ba
L\rightarrow \mbox{Hol}(L,\nabla) \sim 1 + \nabla
\label{heu2}
\ea
due to the loops $L$'s relation to the holonomy map. From (\ref{heu1}) and
(\ref{heu2}) we obtain the non-vanishing commutator
\[
[D,L]\sim [\frac{\d}{\d\nabla},\nabla] \not = 0\;,
\]
which shows a resemblance to a commutation relation
of canonical conjugate variables. 
This means that the Dirac operator is intimately linked to the quantization of
the functional space of connections.

\section{Conclusion and outlook}
\label{sec4}

Noncommutative geometry provides us with an exciting interpretation of the
standard model as a gravitational theory. This formulation has many appealing
features but fails to offer a quantization procedure compatible with the
framework. In the introduction we argued that this problem is inevitable
since a quantization procedure within this form would necessarily include
quantum gravity. Loop Quantum
Gravity provides us with ideas on background independent quantization of
gravity but
lacks, on the other hand, any notion of unification. 
Here we suggest that there
might exist an intersection of the two: 
we study a noncommutative algebra of space-time holonomy loops which is interpreted as
an algebra of functions over a space of connections. The space is described
as a projective system of certain manifolds (Lie-groups) on which spectral triples are
constructed. The whole construction is countable up to (extended) diffeomorphisms. The
inner product in the emerging Hilbert space is interpreted as a functional
integral. Also, the Dirac operator resembles a functional derivation on the
space of connections. Finally, we provide heuristic arguments that a classical
limit might contain an almost commutative geometry which forms the basis of
Yang-Mills-Higgs models.

Many open issues remain. Let us here mention the most important points:
First of all, it is still unclear whether a consistent embedding of Hilbert
spaces exist. In \cite{Aastrup:2005yk} we presented a consistent construction
for non-intersecting loops. For intersecting loops, however, the problem
remains how to construct a metric which is compatible with all projections
between graphs.

Next, it would be natural to consider a construction for non-compact groups,
preferably $SO(3,1)$. This, however, presents difficulties involving both
embedding problems and problems of constructing a spectral triple on
non-compact spaces \cite{Rennie1}.

Finally, it remains to clarify whether the emergent construction satisfy basic requirements of a spectral triple. In particular, we find that the
Dirac operator, in the case of non-intersecting loops, has infinite
dimensional eigenspaces. This points in the direction of semifinite spectral
triples \cite{Rennie2}.

Assuming these difficulties can be resolved, an interesting question to address is that of a semiclassical limit. One
may speculate whether perturbations around a classical limit can generate some
kind of 'quantization' of the fields which presumable emerge.

\section{Acknowledgment}

It is a pleasure to thank Victor Gayral, Troels Harmark, Ryszard Nest, Adam Rennie and Raimar Wulkenhaar for discussions
and comments. We thank Raimar Wulkenhaar and the mathematics institute at M\"unster University for
hospitality during a visit.


\begin{thebibliography}{99}




\bibitem{ConnesBook}
A.~Connes,
``Noncommutative Geometry,'' Academic Press, 1994.

\bibitem{Aastrup:2005yk}
  J.~Aastrup and J.~M.~Grimstrup,
  ``Spectral triples of holonomy loops,''
  arXiv:hep-th/0503246.

\bibitem{Connes:1996gi}
A.~Connes,
``Gravity coupled with matter and the foundation of non-commutative
geometry,''
Commun.\ Math.\ Phys.\  {\bf 182} (1996) 155
[arXiv:hep-th/9603053].

\bibitem{Connes:1990qp}
A.~Connes and J.~Lott,
``Particle Models And Noncommutative Geometry (Expanded Version),''
Nucl.\ Phys.\ Proc.\ Suppl.\  {\bf 18B} (1991) 29.

\bibitem{Chamseddine:1991qh}
A.~H.~Chamseddine and A.~Connes,
``Universal formula for noncommutative geometry actions: Unification of
gravity and the standard model,''
Phys.\ Rev.\ Lett.\  {\bf 77} (1996) 4868.
 
\bibitem{Chamseddine:1996rw}
A.~H.~Chamseddine and A.~Connes,
``A universal action formula,''
[arXiv:hep-th/9606056].

 
\bibitem{Jureit:2005tw}
  J.~H.~Jureit, T.~Schucker and C.~Stephan,
  ``On a classification of irreducible almost commutative geometries. III,''
  J.\ Math.\ Phys.\  {\bf 46} (2005) 072303
  [arXiv:hep-th/0503190].

\bibitem{Schucker:2005aa}
  T.~Schucker,
  ``Krajewski diagrams and spin lifts,''
  arXiv:hep-th/0501181.

\bibitem{Jureit:2005ay}
  J.~H.~Jureit and C.~A.~Stephan,
  ``On a classification of irreducible almost commutative geometries, a second
  helping,''
  J.\ Math.\ Phys.\  {\bf 46} (2005) 043512
  [arXiv:hep-th/0501134].

\bibitem{Iochum:2003xy}
  B.~Iochum, T.~Schucker and C.~Stephan,
  ``On a classification of irreducible almost commutative geometries,''
  J.\ Math.\ Phys.\  {\bf 45} (2004) 5003
  [arXiv:hep-th/0312276].


\bibitem{Ashtekar:2004eh}
  A.~Ashtekar and J.~Lewandowski,
  ``Background independent quantum gravity: A status report,''
  Class.\ Quant.\ Grav.\  {\bf 21} (2004) R53
  [arXiv:gr-qc/0404018].


\bibitem{Rovelli:2004tv}
  C.~Rovelli,
  ``Quantum gravity,''
Cambridge, UK: Univ. Pr. (2004) 455 p.


\bibitem{Smolin:2004sx}
  L.~Smolin,
  ``An invitation to loop quantum gravity,''
  arXiv:hep-th/0408048.


\bibitem{Fairbairn:2004qe}
  W.~Fairbairn and C.~Rovelli,
  ``Separable Hilbert space in loop quantum gravity,''
  J.\ Math.\ Phys.\  {\bf 45} (2004) 2802
  [arXiv:gr-qc/0403047].



\bibitem{Ashtekar:1993wf}
A.~Ashtekar and J.~Lewandowski,
``Representation theory of analytic holonomy C* algebras,''
[arXiv:gr-qc/9311010].




\bibitem{Marolf:1994cj}
D.~Marolf and J.~M.~Mourao,
``On the support of the Ashtekar-Lewandowski measure,''
Commun.\ Math.\ Phys.\  {\bf 170} (1995) 583
[arXiv:hep-th/9403112].




\bibitem{Ashtekar:1994wa}
  A.~Ashtekar and J.~Lewandowski,
 ``Differential geometry on the space of connections via graphs and projective
  limits,''
  J.\ Geom.\ Phys.\  {\bf 17} (1995) 191
  [arXiv:hep-th/9412073].


\bibitem{Rennie1}
A.~Rennie, ``Smoothness and locality for nonunital spectral triples'', {\it
  K-Theory} {\bf 28} (2003) 127.

\bibitem{Rennie2}
A.~L.~Carey, J.~Phillips, A.~Rennie and F.~A.~Sukochev,
``The Local Index Formula in Semifinite von Neumann Algebras I: Spectral Flow''
 [arXiv:math.OA/0411019]
  










\end{thebibliography}
\end{document}